\documentclass[a4paper]{jpconf}
\usepackage{graphicx}

\newcommand{\ttbar}{\ensuremath{\mathrm{t\overline{t}}}}
\newcommand{\dzero}{\ensuremath{\mathrm{D^0}}}
\newcommand{\dpm}{\ensuremath{\mathrm{D^{\pm}}}}
\newcommand{\bzero}{\ensuremath{\mathrm{B^0}}}
\newcommand{\bpm}{\ensuremath{\mathrm{B^{\pm}}}}
\newcommand{\jpsi}{\ensuremath{\mathrm{J/\Psi}}}
\newcommand{\ztwostar}{\ensuremath{\mathrm{Z2^*}}}
\renewcommand{\pt}{\ensuremath{p_{T}}}

\begin{document}
\title{Underlying Event and B-Hadron Decays in \ttbar{} Events}

\author{Benjamin Stieger \\
{\it on behalf of the CMS collaboration\footnote{Presented at the 7\textsuperscript{th} International Workshop on Top Quark Physics, 29\textsuperscript{th} September to 3\textsuperscript{rd} October 2014 -- Cannes, France}}}

\address{CERN, PH department, CH-1211 Geneva-23, Switzerland}

\ead{stiegerb@cern.ch}

\begin{abstract}
We present exploratory studies of the underlying event activity and of fragmentation and hadronization of b quarks using \ttbar{} candidate events in proton-proton collision data acquired by the CMS experiment.
We reconstruct charm mesons in fully charged decay channels from the reconstructed tracks associated with the hadronization of b quarks from the top decay, and study their kinematics relative to the mother jet.
A good agreement is found using \textsc{MadGraph} plus the \textsc{Pythia}~6 Tune \ztwostar{} simulation.
The effects predicted by alternative settings and generators for the characterization of the underlying event are also explored.
These results are expected to contribute in the future to more precise measurements in the top quark sector in particular of the top quark mass by either constraining systematic uncertainties related to the modeling of the underlying event in \ttbar{} events or by paving the way for alternative mass measurement methods.
\end{abstract}

\section{Introduction}
Measurements of the top quark mass and other properties generally rely on relating the observed final states and parton-level kinematics.
However, due to the fact that the top is a colored particle, its decay products necessarily have to reconnect to other colored states in the event to form color-neutral stable particles.
Moreover, as the top anti-top pair is produced from the splitting of a gluon in hadronic collisions, this color connection cannot happen entirely within the products of the hard interaction, but inevitably involves the remnants of the initial protons.
Therefore the full reconstruction of the initial kinematics may be compromised by both the ability of the jet algorithm to capture the products of the fragmentation of the b and the interplay between the b quark and the underlying event (UE) or other jets produced in the collision.

The large available statistics of \ttbar{} events in the first running period of the Large Hadron Collider (LHC) allows for the first time to study these effects directly at the scale of production and decay of the top quark.
We make use of the high-purity and high-statistics sample collected at $\sqrt{s}=8~\mathrm{TeV}$ to compare the predictions of relevant observables from different Monte-Carlo (MC) generators with different tunes, with the ultimate goal of reducing systematic uncertainties by using models that better reflect our knowledge of these effects.
The used sample corresponds to an integrated luminosity of $19.7\pm0.5~\mathrm{fb^{-1}}$.
Two results are reported in this manuscript: the characterization of underlying event properties in \ttbar, and the study of kinematic properties of reconstructed charm mesons within the top decay products~\cite{top-13-007}.

\section{Underlying Event Studies}
We define the underlying event in dileptonic \ttbar{} events as any hadronic activity that cannot be attributed to the particles arising from the hard scattering, i.e.\ the decay products of the \ttbar{} system.
The hadronization of initial- and final-state radiation (ISR/FSR) is considered as part of the UE unless the particles are clustered in the reconstructed b-jet candidates.
We select top pair events in the dileptonic channel by requiring two isolated high-\pt{} leptons (electrons or muons), compatible with the primary interaction vertex.
Furthermore we require at least two b-tagged jets in the event.
With this event selection, we expect a purity of about 83\% \ttbar{} events in the same-flavor channels, and about 96\% in the e$\mu$ channel.

We use the charged candidates of the CMS particle flow algorithm~\cite{PF} to define observables related to the underlying event.
Candidates clustered in the b-jets or associated to the leptons from the top decay are discarded.
The remaining candidates are selected to have $\pt>500~\mathrm{MeV}$ and $|\eta|<2.1$ and are required to be consistent with the primary interaction vertex to suppress contributions from concurrent proton-proton interactions (pileup).

We study the charged multiplicity, transverse flux, and average flux per charged particle in various regions of phase space defined by the number of additional jets and by the azimuthal direction relative to the total momentum of the \ttbar{} system.
Three regions in the transverse plane are considered, using the relative $\phi$-angle to the total \ttbar{} momentum: ``toward'' if $\Delta\phi<60^\circ$, ``transverse'' if $\Delta\phi$ is within $60-120^\circ$, and ``away'' if $\Delta\phi>120^\circ$.

\begin{figure}[!htb]
  \begin{center}
	\includegraphics[width=0.32\linewidth]{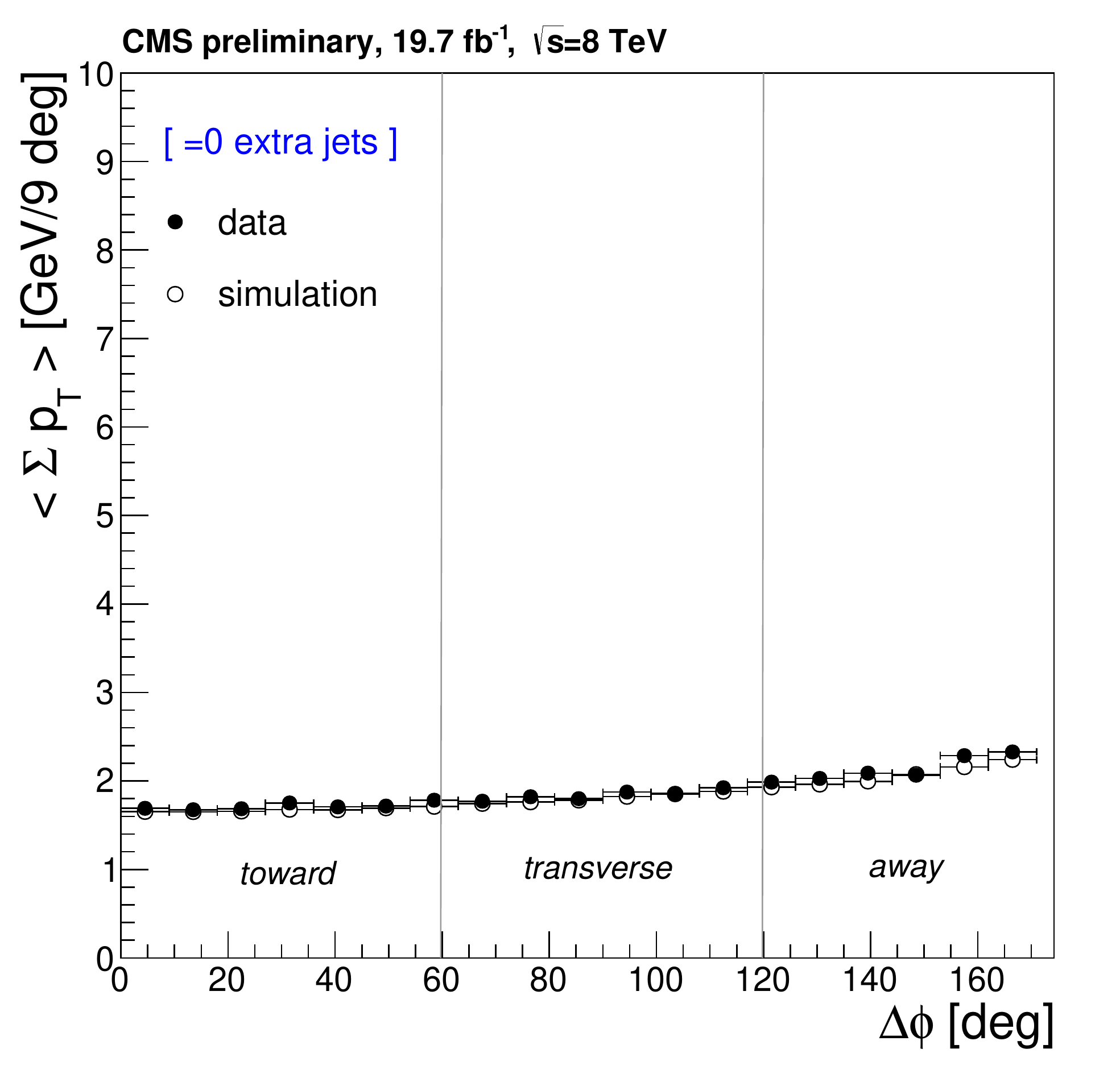}
	\includegraphics[width=0.32\linewidth]{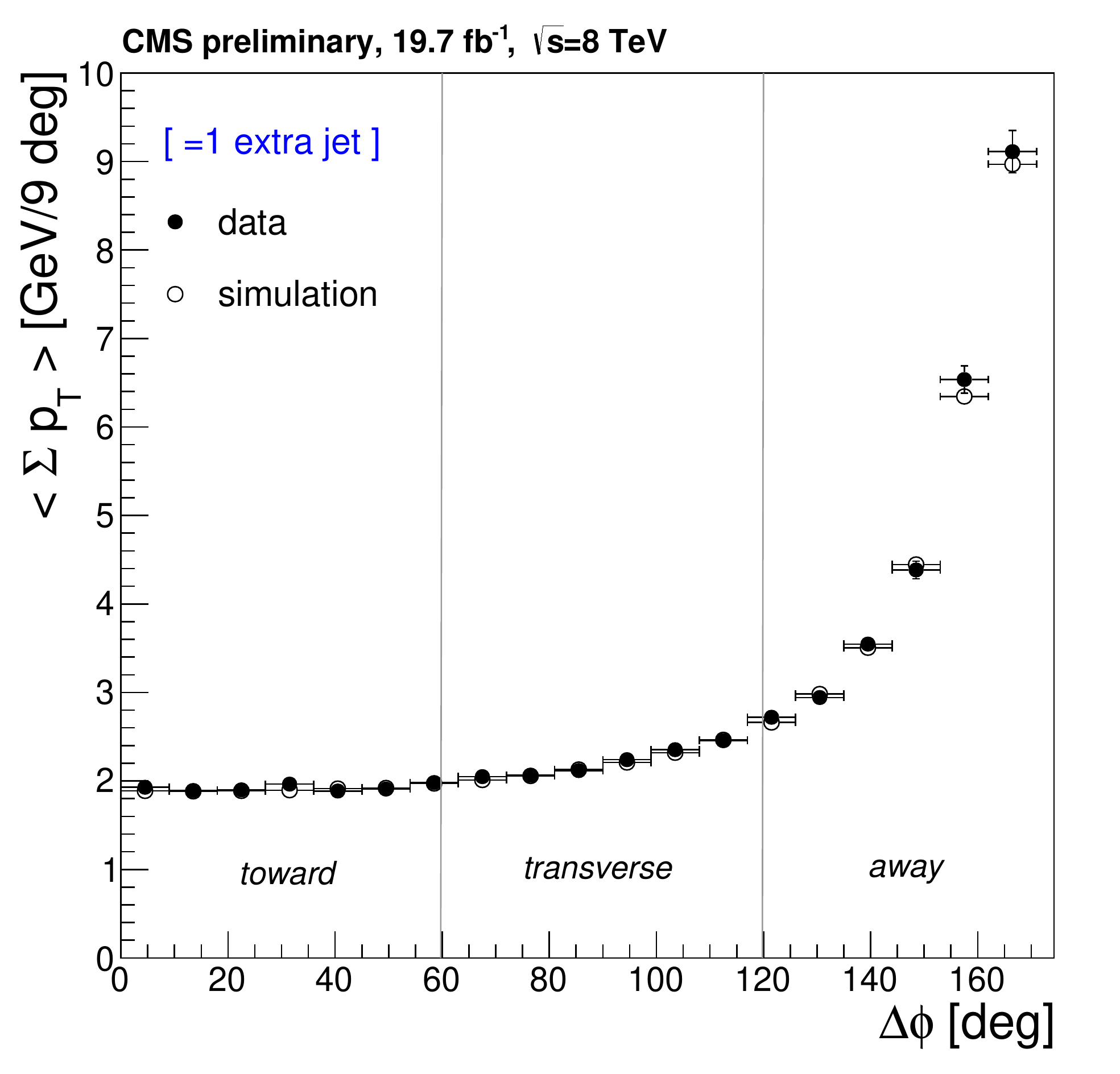}
	\includegraphics[width=0.32\linewidth]{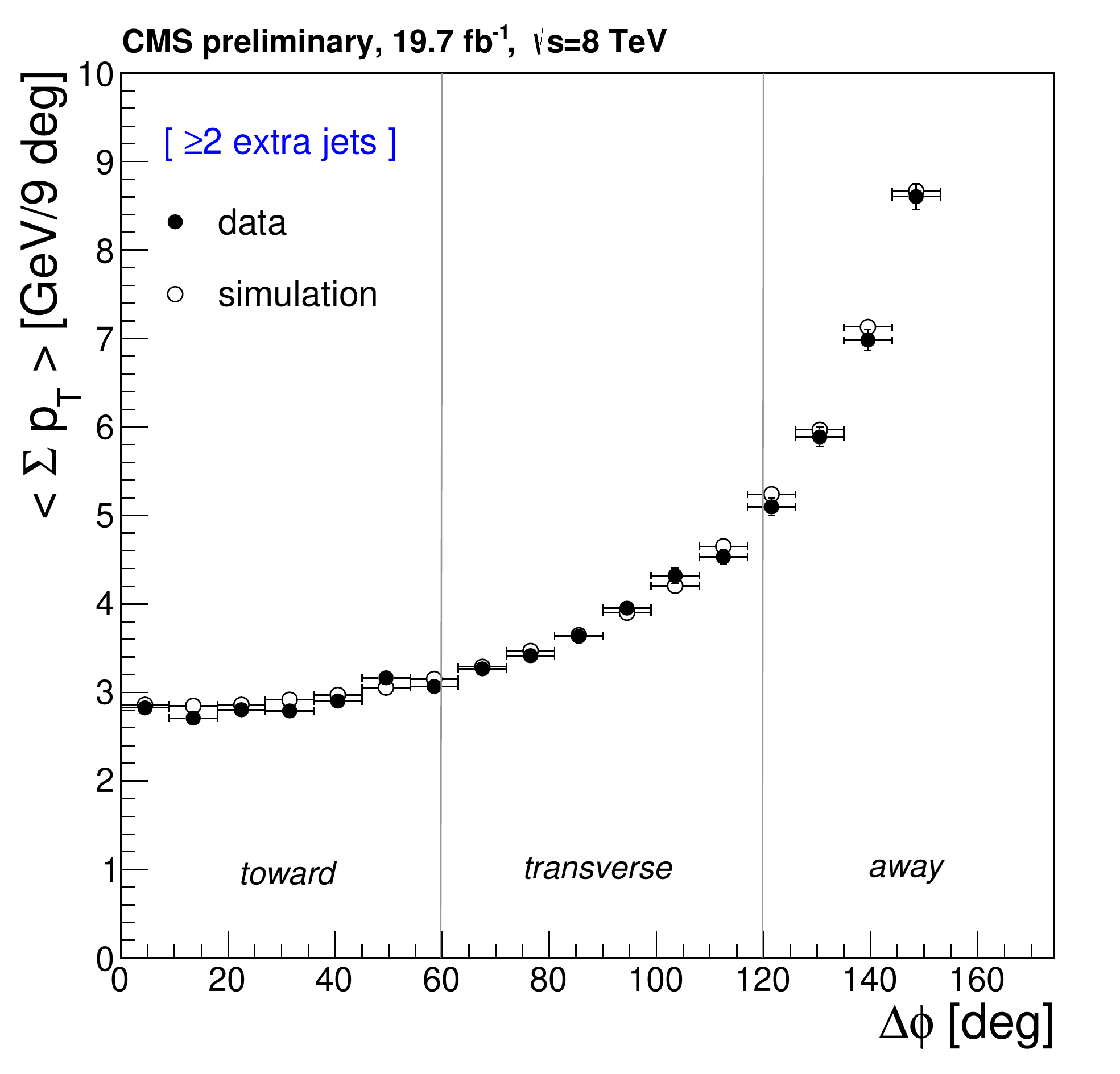}
  \end{center}
  \caption{Average transverse flux in bins of azimuthal distance to the transverse direction of the \ttbar{} system for events with no extra jets (left), exactly one extra jet (middle), and two or more extra jets (right). Data is compared with the prediction from \textsc{MadGraph}+\textsc{Pythia}~6~\ztwostar.\label{fig:ue1}}
\end{figure}

Figure~\ref{fig:ue1} shows the integrated transverse flux profiled against $\Delta\phi$, for events with no extra jets, exactly one extra jet, and two or more extra jets.
In events with extra jets present, a sharp rise of the flux in the away region is observed, consistent with the expectation of the hadronic recoil to the \ttbar{} system.
The observed data distributions are described well by \textsc{MadGraph}+\textsc{Pythia}~6 \ztwostar{}\cite{Z2star}.

\begin{figure}[!htb]
  \begin{center}
	\includegraphics[width=0.40\linewidth]{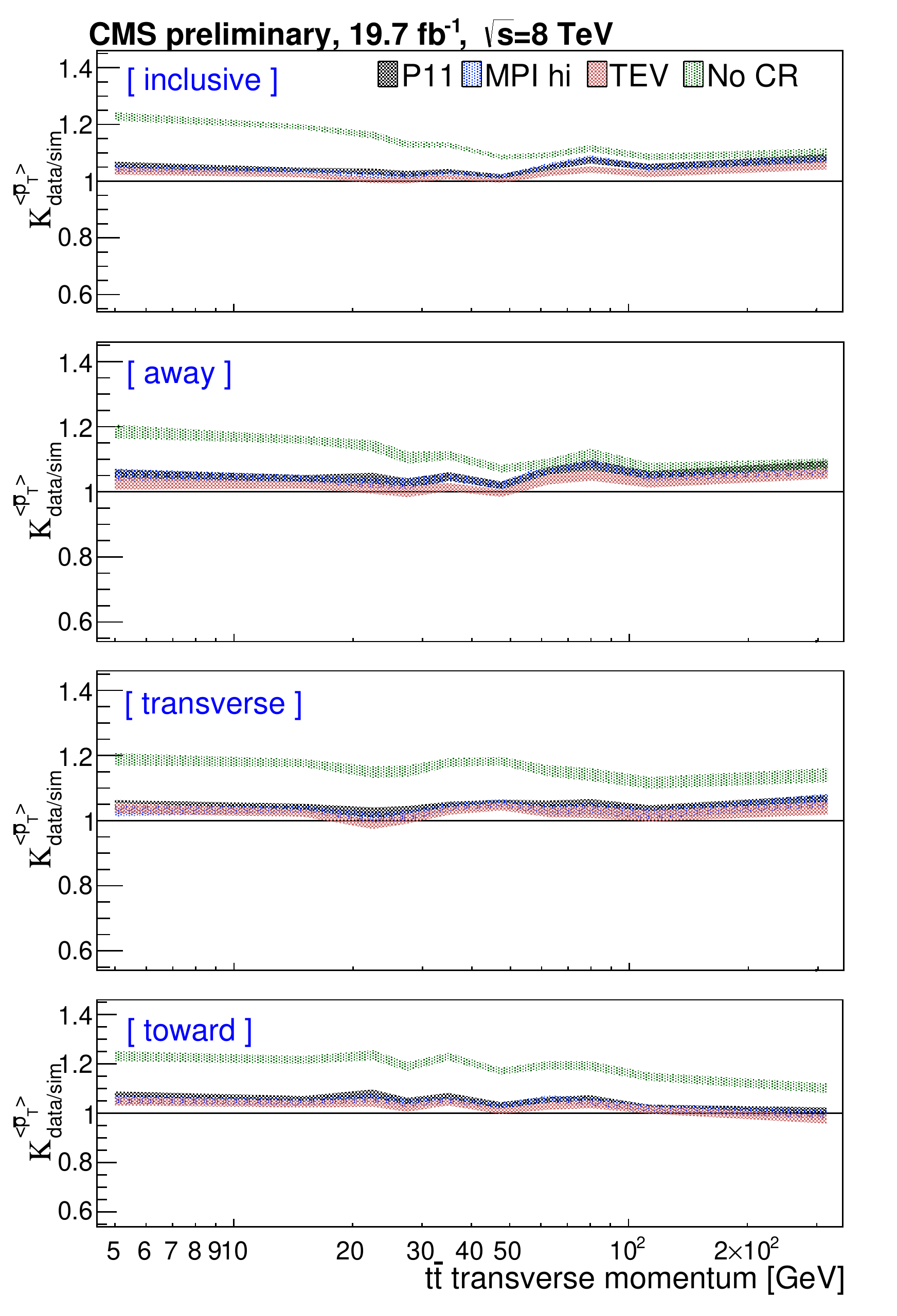}
	\includegraphics[width=0.40\linewidth]{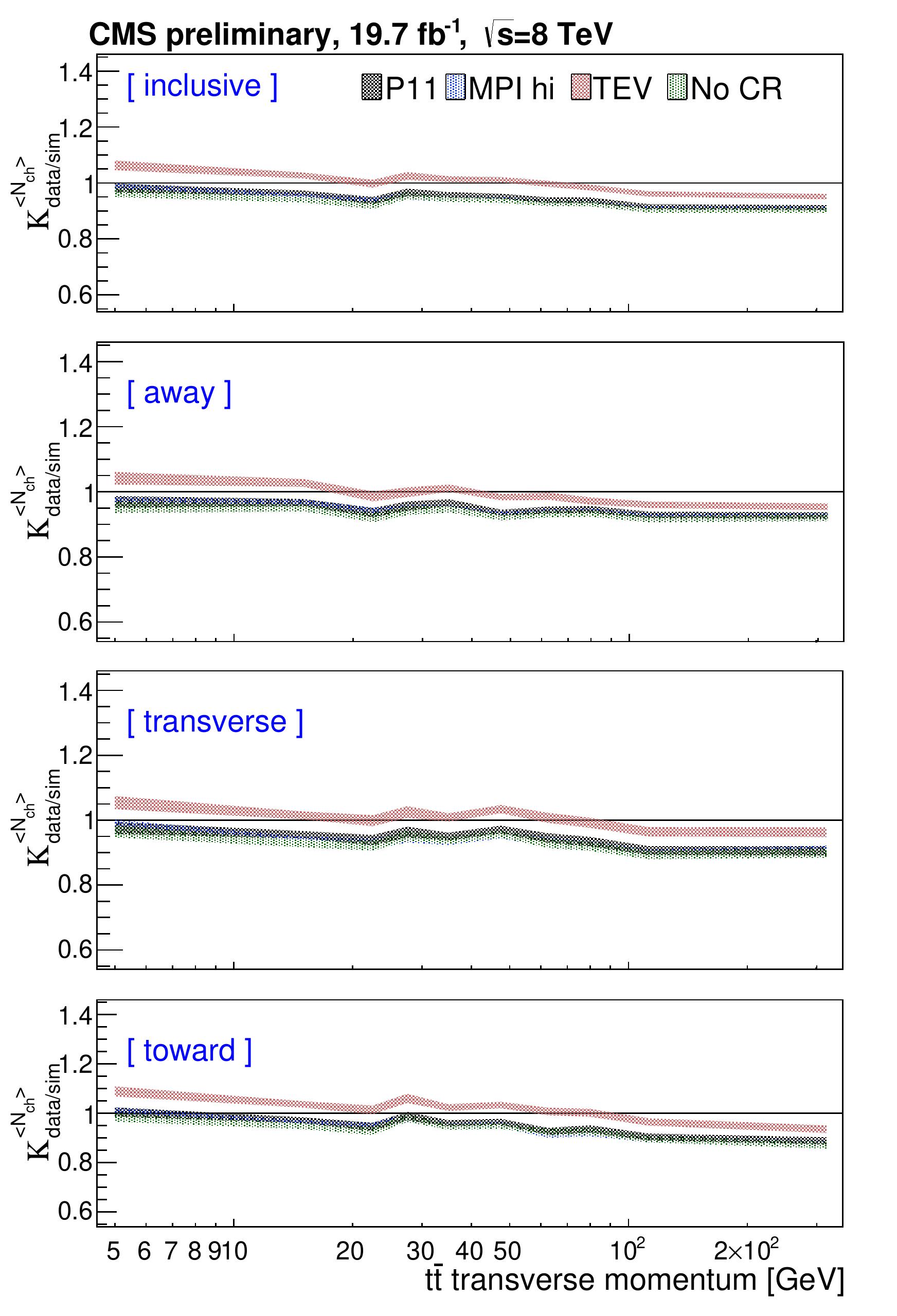}
  \end{center}
  \caption{Ratio of data/simulation for average particle \pt{} (left) and average charge multiplicity (right), as a function of the \ttbar{} transverse momentum, in the regions away from, transverse to, and towards the \ttbar{} momentum, as well as the inclusive.\label{fig:ue2}}
\end{figure}

Figure~\ref{fig:ue2} shows the ratio of data to simulation for the average \pt{} per particle, and the average number of particles as a function of the \ttbar{} transverse momentum, in the away, transverse, toward, and inclusive regions of the transverse plane.
Four different \textsc{Pythia}~6 tunes are compared: nominal Perugia '11 (P11); enhanced multi-parton interactions (MPI hi); tuned to Tevatron data (TEV); and suppressed color-reconnection (No CR)\cite{P11}.
Compared to the data, all tunes tend to predict a smaller number of particles with larger average \pt{}, in particular in events with a significant amount of transverse momentum in the \ttbar{} system, while the overall flux agrees well.
The variation with suppressed color-reconnection is strongly disfavored by the data.

\section{B-Hadron Decays}
One of the major sources of systematic uncertainties in top mass measurements concerns the modeling of the hadronization and fragmentation of b-quarks. 
We perform the first step towards an in-situ measurement of b-fragmentation in \ttbar{} events by reconstructing charm mesons within the decay products of the b-jets associated with the top decays~\cite{top-13-007}.
We study the kinematics of the charm meson candidates relative to the mother jet and compare various generator setups and underlying event tunes.

We select events in the dilepton channels as for the underlying event study, but also consider the single lepton channels, by requiring a hard lepton and at least four jets.
In the selected events, we then consider the two jets with the highest discriminator value from the constrained secondary vertex (CSV) b-tagging algorithm, and attempt to reconstruct charm meson candidates within the charged particle flow candidates associated with them.

Three mesons are considered: \dzero{} in the $\mathrm{K}^\pm\pi^\mp$ channel, \dpm{} in the $\mathrm{K}^\mp\pi^\pm\pi^\pm$ channel, and \jpsi{} in the $\mu^+\mu^-$ channel.
In order to tag the charm flavor, and correctly assign the kaon and pion mass hypotheses to the tracks in the case of the \dzero, we target the semi-leptonic decay of \bzero{} and \bpm{} and use the charge of a soft muon as the flavor tag.

\begin{figure}[!htb]
  \begin{center}
	\includegraphics[width=0.32\linewidth]{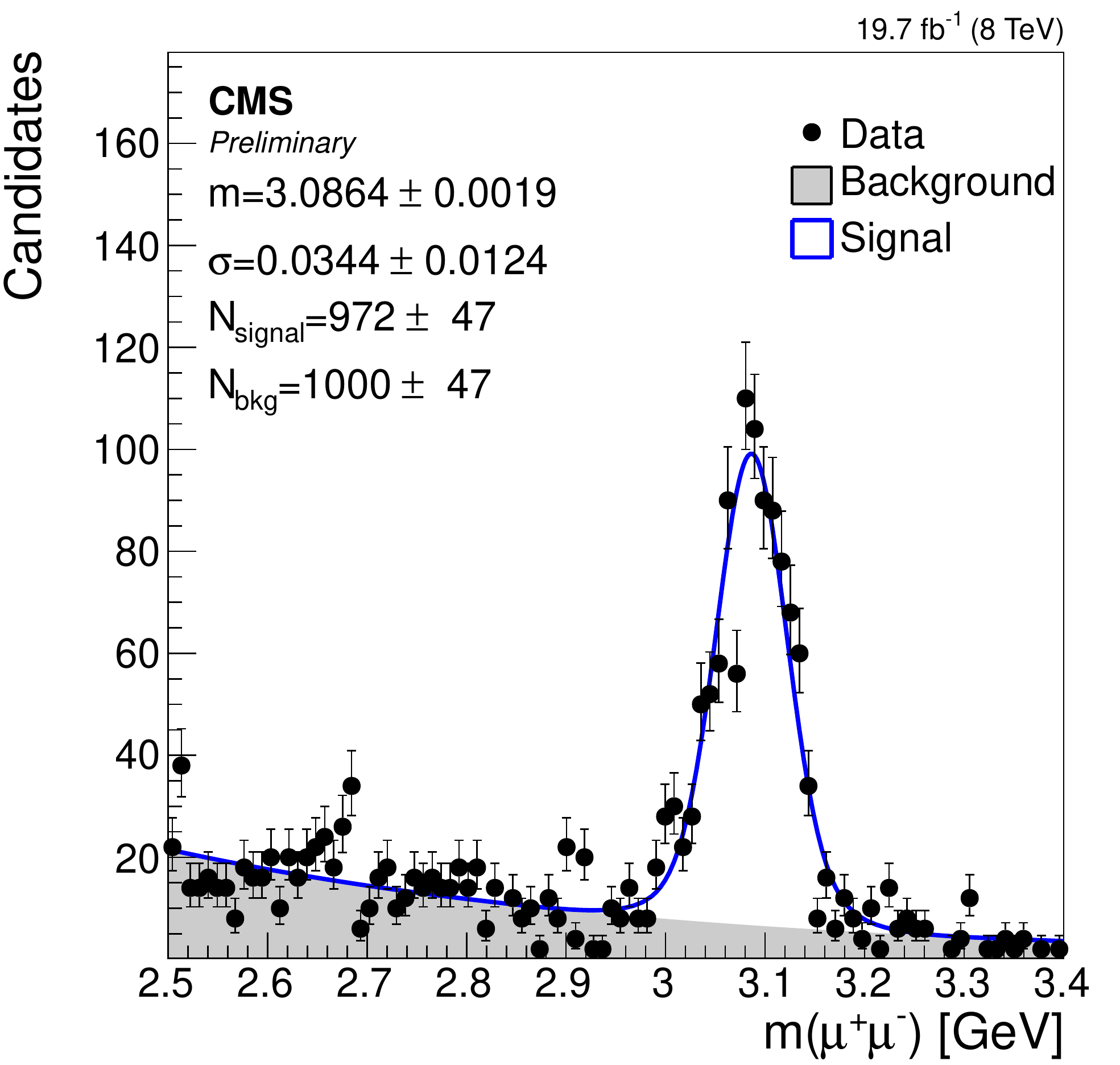}
	\includegraphics[width=0.32\linewidth]{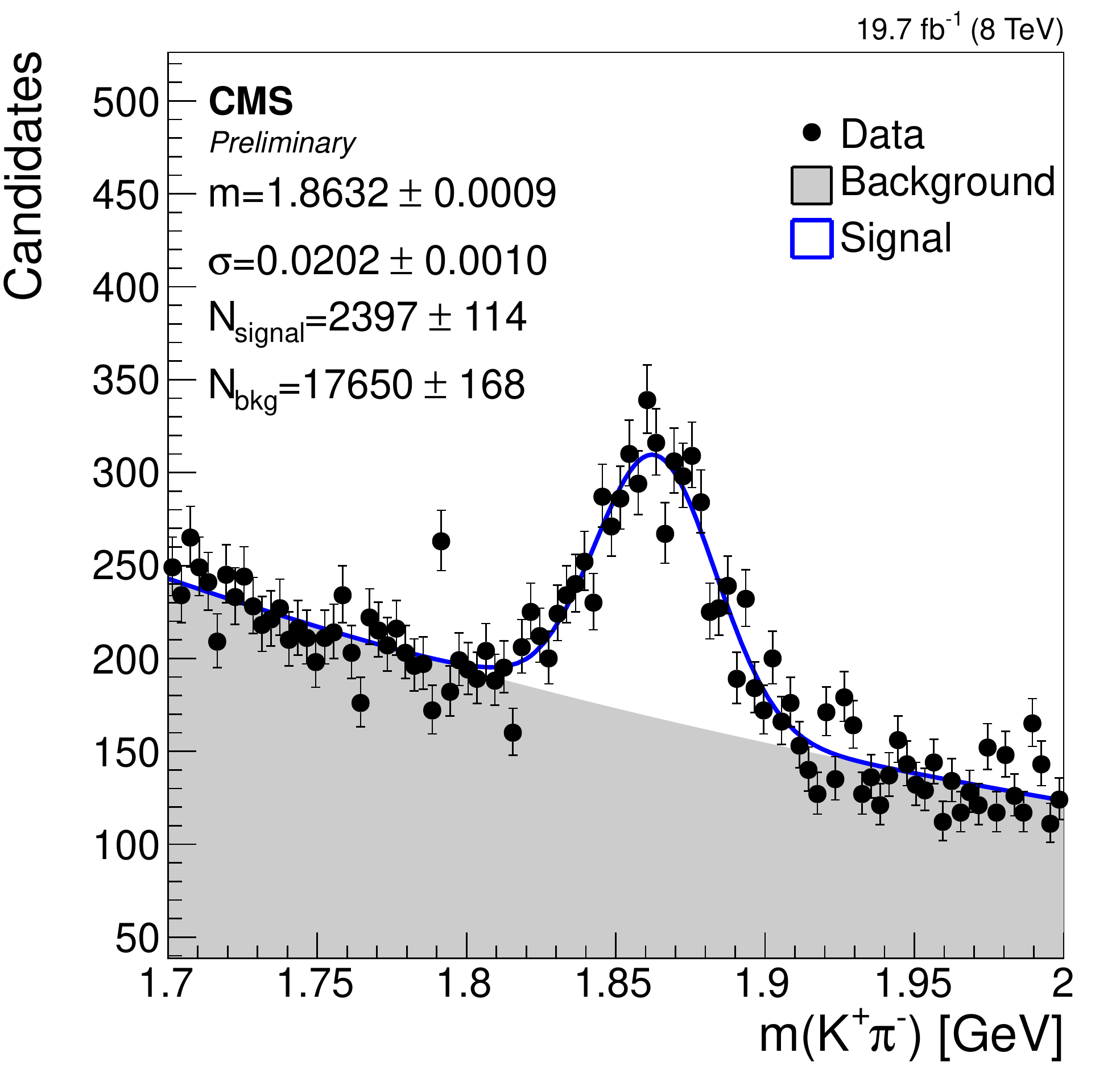}
	\includegraphics[width=0.32\linewidth]{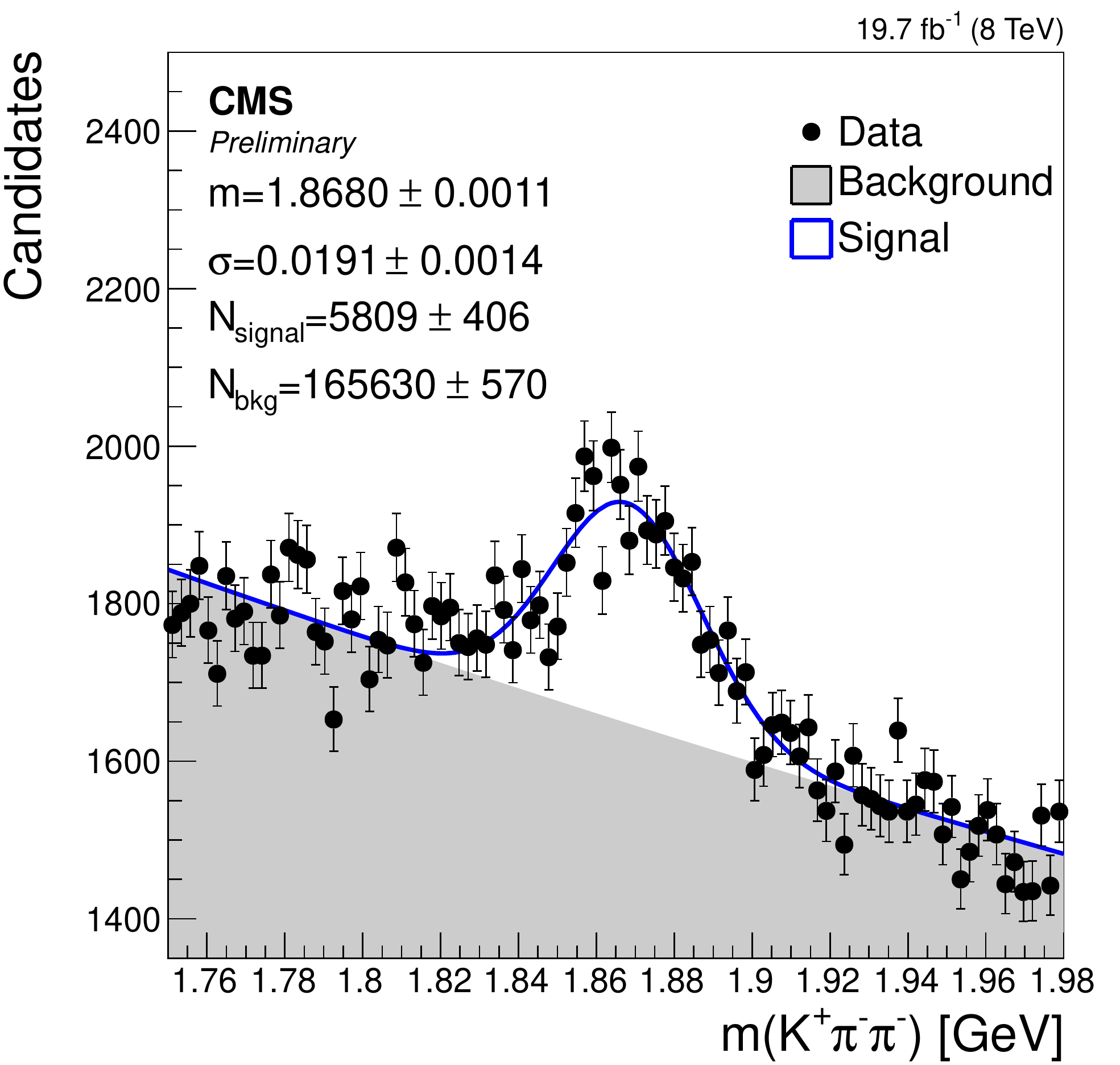}
  \end{center}
  \caption{Invariant masses of reconstructed \jpsi{} (left), \dzero{} (middle), and \dpm{} (right) candidates, reconstructed within the decay products of top quarks.\label{fig:frag1}}
\end{figure}

The reconstructed invariant masses of charm meson candidates can be seen in Fig.~\ref{fig:frag1}.
We obtain samples of roughly 1'000 \jpsi{}, about 2'400 $\mu$\dzero{}, and about 5'800 \dpm{} candidates.

\begin{figure}[!htb]
  \begin{center}
	\includegraphics[width=0.40\linewidth]{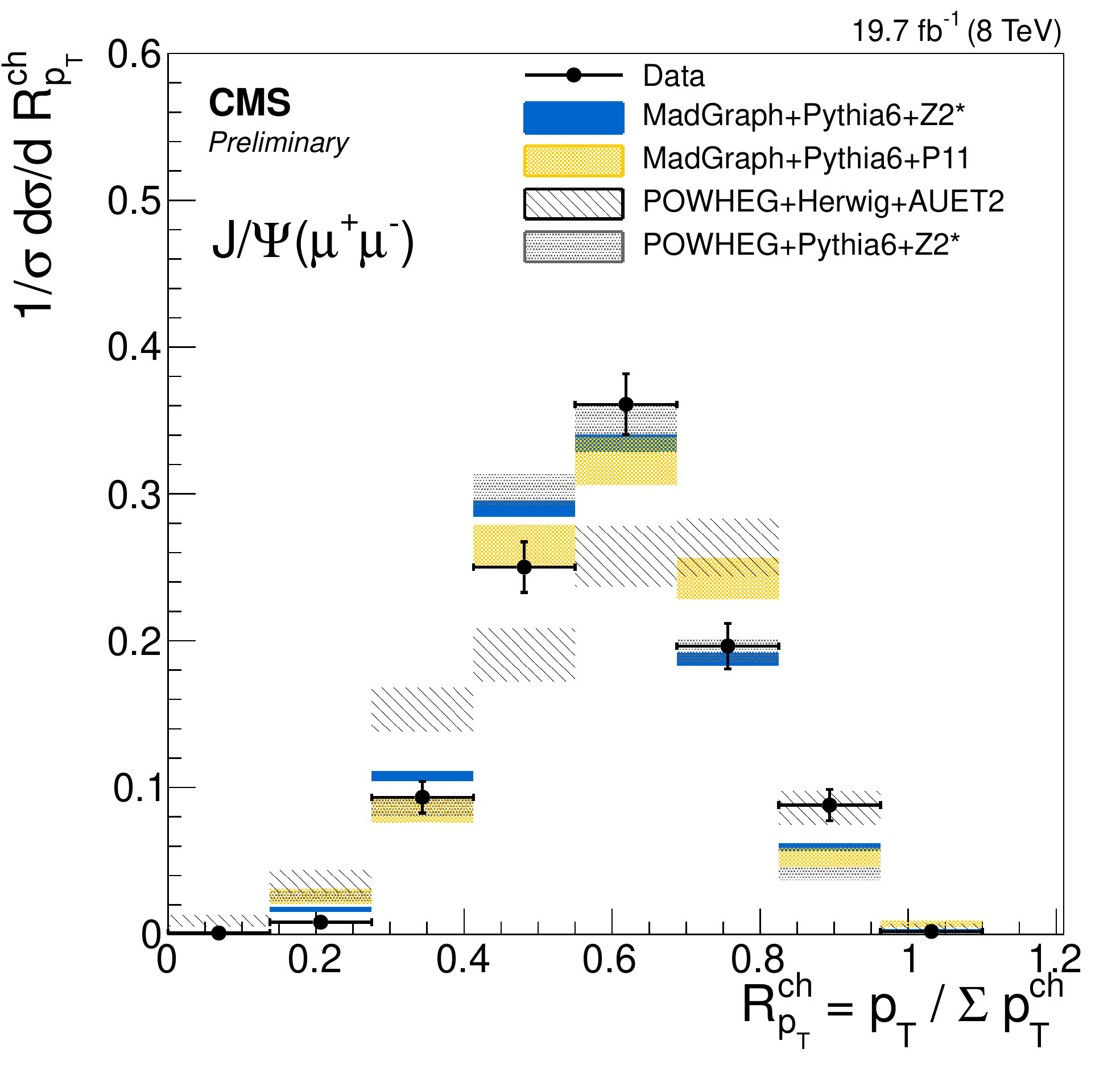}
	\includegraphics[width=0.40\linewidth]{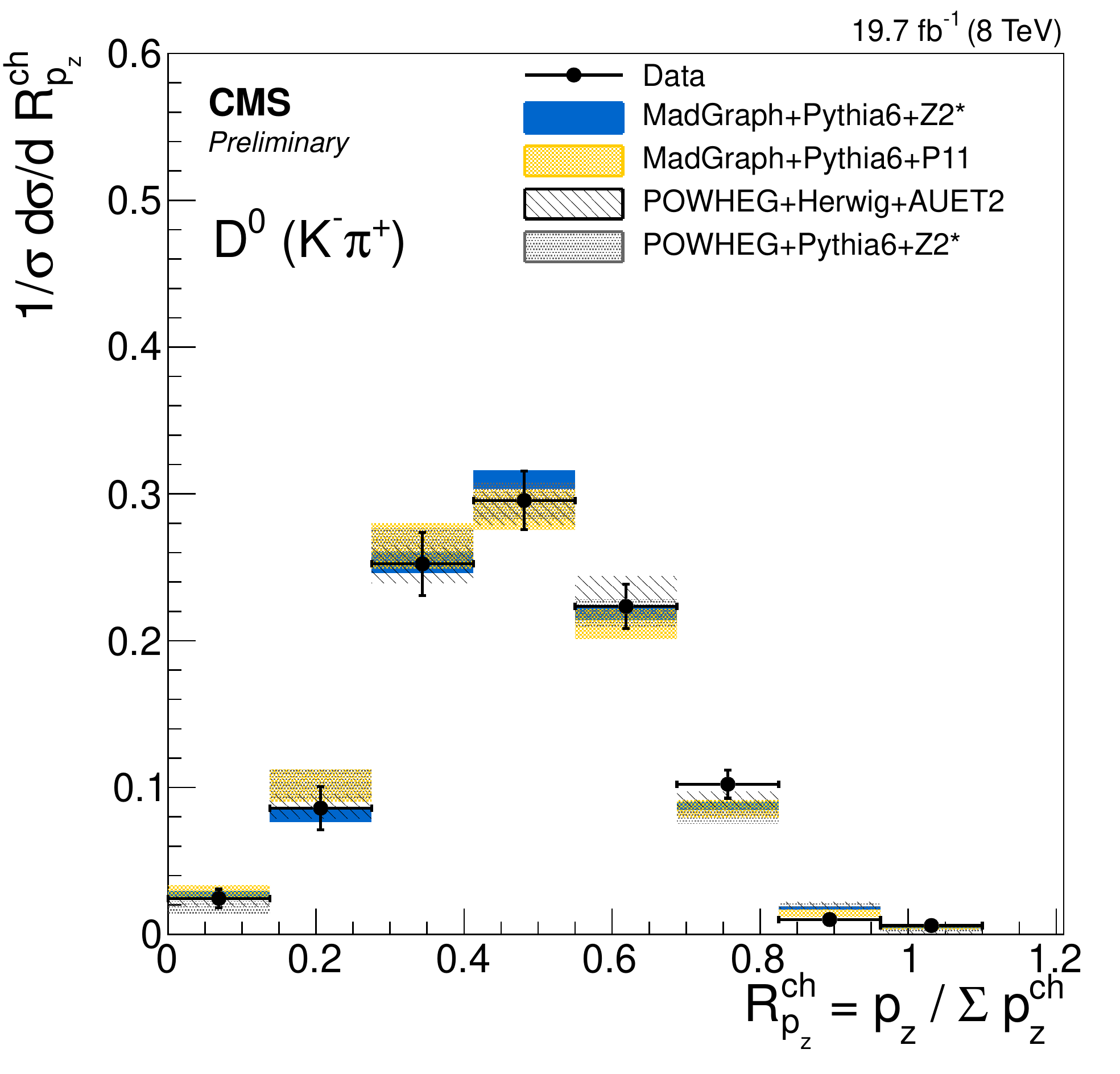}
  \end{center}
  \caption{Ratio of candidate \pt{} and charged \pt{} of the mother jet for \jpsi{} (left) and \dzero{} mesons (right). Observed data is predicted to several Monte-Carlo generator setups.\label{fig:frag2}}
\end{figure}

Using the distribution in the reconstructed invariant mass, we can perform a fit and use the resulting probability density functions to separate the kinematics of charm candidates and the combinatorial background with the sPlot technique~\cite{splot}.
We consider a range of kinematic observables of the charm candidates: their transverse momentum (\pt), the ratio of transverse and longitudinal momentum of the candidates with respect to the mother jet's transverse and longitudinal momenta, the \pt{} of the candidate relative to the mother jet system, and the angular distance between candidate and the jet.
As an example, figure~\ref{fig:frag2} shows the ratio of candidate \pt{} to the \pt{} of the mother jet, calculated using only the associated charged tracks, for \jpsi{} and $\mu$\dzero.
We compare the data to predictions of \textsc{MadGraph} with two different \textsc{Pythia}~6 tunes, and POWHEG interfaced with Herwig and the ATLAS underlying event tune (AUET), as well as \textsc{Pythia}~6 with the \ztwostar{} tune.
In the case of \jpsi{} the Herwig prediction is strongly disfavored by the data, whereas the default \textsc{Pythia}6+\ztwostar{} describes the data fairly well in all distributions.

\end{document}